\documentstyle[prl,aps,epsfig]{revtex}

\newcommand \be{\begin{equation}}
\newcommand \ee{\end{equation}}
\newcommand \beq{\begin{eqnarray}}
\newcommand \eeq{\end{eqnarray}}
\newcommand{\set}[2]{\newcommand{#1}{#2}}
\set{\pa}{\partial \over \partial}
\set{\leftvector}{\stackrel{\leftarrow}{\partial}}
\set{\rightvector}{\stackrel{\rightarrow}{\partial}}

\begin{document}
\twocolumn [ \hsize\textwidth\columnwidth\hsize
             \csname @twocolumnfalse\endcsname
\author{C. Simenel, Ph. Chomaz, G. de France}
\title{Quantum calculation of dipole excitation in fusion reactions}
\date{\today}
\address{G.A.N.I.L., B.P. 55027,F-14076 Caen Cedex 5, France.}
\maketitle

\begin{abstract}
The excitation of the giant dipole resonance induced by fusion reaction is studied with N/Z
asymmetry in the entrance channel. The Time Dependant Hartree Fock solution
 exhibits a strong dipole vibration which can be associated to a giant 
 vibration along the main axis of the deformed compound nucleus. This dipole motion appears 
 to be non linearly coupled to the shape oscillation leading to a strong modulation of
 its frequency. These phenomenons can be detected in the gamma-ray emission from hot 
 compound nuclei.
\end{abstract}

]

\pacs{Pre-equilibrium; Dipole
Vibration;
N/Z
Asymmetry;
TDHF;
Fusion;
Entrance
Channel; Isospin}

  Ordered collective motions are a
 general property of mesoscopic systems. In
metallic clusters, electron vibrations are plasmon excitations. In
atomic nuclei, oscillations of protons against neutrons generate giant
dipole resonances \cite{bal,gol}. The general way to excite such modes is to
use rapidly varying electromagnetic fields associated with photons or
generated by fast electrically charged particles. The collective vibrations
can also be thermally excited as it was clearly demonstrated in the studies
of the $\gamma $ - emission from hot nuclei\cite{Brink,Sno86,Gaa88,Bra89}. It has
been recently proposed that fusion reactions with N/Z asymmetric nuclei may
lead to the excitation of a dipole mode because of the presence of a net
dipole moment in the entrance channel\cite{cho2,DiToro,bar}.
The first experimental indications on the possible existence of such new
phenomenon have been reported in \cite{fli} for fusion reactions and in 
\cite{san} for deep inelastic collisions. However, the real nature of such a
vibration is still unclear both from the experimental and the
theoretical point of view. In particular only semiclassical approaches or
schematic models have been used to infer the properties of the generated
dipole mode. 

In this letter, we present the first quantum calculation of pre-equilibrium giant collective vibrations
 using the time dependent Hartree Fock (TDHF) approach \cite
{har,foc,vau,bon,neg}. TDHF corresponds to an independent propagation of
each single particle wave function in the mean field generated by the
ensemble of particles. It does not incorporate the dissipation due to
two-body interaction \cite{gon,won,lac}, but takes into account one body mechanisms
such as Landau spreading and evaporation damping \cite{cho-Landau}. The quantal 
nature of the single particle dynamics is explicitely preserved, which
is crucial at low energy both because of shell effects and of the wave
dynamics. Moreover TDHF is a strongly non linear theory. Hence it can exhibit new couplings 
between collective modes.

  \smallskip In the  time
 dependent Hartree-Fock (TDHF) approach, the
evolution of the single particle density matrix $\rho
(t)=\sum_{n=1}^{N}\left| \varphi _{n}\right\rangle \left\langle \varphi
_{n}\right| $ is determined by a Liouville equation,

\begin{equation}
i\hbar \frac{\partial }{\partial t}\rho -[h(\rho ),\rho ]={\rm 0}
\end{equation}
where $h(\rho )$ is the mean-field Hamiltonian. We have used the code built 
by P. Bonche and coworkers  with an effective Skyrme mean-field and $SLy_{4}$ parameters  \cite{kim}. 

\begin{figure}[tbph]
\begin{center}
\epsfig{file=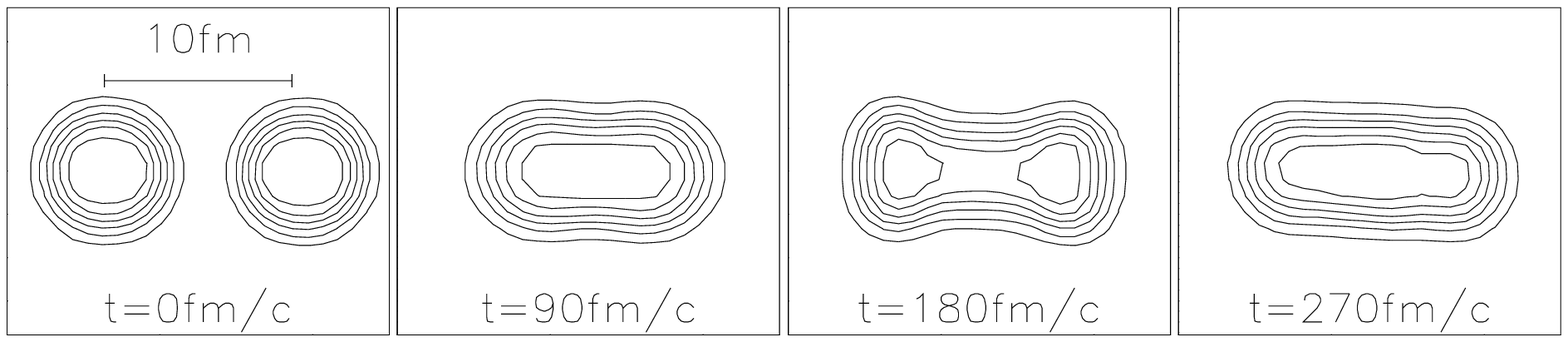,height=2.0cm,width=7.0cm} 
\end{center}
\caption{{{{{{{{\protect\footnotesize {Density plots, projected on the
reaction plane, for the central collision reaction $^{20}O+^{20}Mg$ at 1
MeV/u. Lines corresponds to equidistant values of the density.}}}}}}}}}
\label{fig:density}
\end{figure}

  The effect of the isospin asymmmetry in the entrance channel has
 been first studied in the $^{20}O+^{20}Mg$ fusion reactions at energies close to 
the Coulomb barrier. Strong quantum effects are expected in these mirror-nuclei reactions 
leading to the N=Z $^{40}Ca$ compound system. The density plots obtained in 
the central collisions at a kinetic energy of $1 MeV$ per nucleon  in the center of mass frame
is presented in Fig. $1$. The system  fuses and shows 
quadrupole oscillation around a slowly damped prolate deformation. Since $%
^{20}O$ and $^{20}Mg$ have significantly different N over Z ratios (respectively
1.5 and 0.67), the protons and neutrons centers of mass do not coincide
 at the initial stage 
 of the collision in contrast to what is obtained in the fusion of two  nuclei
with the same N over Z ratio. 

Let us first start with head on head collisions along the $x$ axis.
Adequate observables to study the collective motion induced in this mechanism
 are the dipole moment $Q_{d}$ and its conjugated quantity $P_{d}$.
 $Q_{d}$ is the net distance between protons and neutrons:
 $Q_{d}=\frac{N Z}{A} (X_{p} - X_{n} )$ where $X_{p}$ and 
$X_{n}$ are the proton and neutron's centers of mass coordinates.
 Similarly, $P_{d}=\frac{A}{2 N Z} (P_{p} - P_{n})$
 where $P_{p}=\sum_{p} p_p$ and $P_{n}=\sum_{n} p_n$ are the moments 
of protons and neutrons. In Fig. $2$ is plotted $Q_{d}$ as a function of $P_{d}$. 
As time goes on, we observe a spiral 
in the collective phase space which signals the presence
of a damped collective vibration. Indeed, it originates from oscillations
in phase quadrature of the two conjugated dipole variables.

  In order to associate the observed vibration with the giant
dipole resonance (GDR) in the composite nucleus we must study the collective
vibration of a $^{40}Ca$ nucleus. 
The GDR associated to the dipole oscillation of a $^{40}Ca$ ground state has
 been excited by an isovector dipole field and followed using the TDHF approach. The resulting $Q_{d}$ moment is 
plotted as a function of time in Fig. $3-c$. The period of the observed oscillations is around 
$80$ fm/c which corresponds to an energy $E=15.5 MeV$ in good agreement with skyrme 
RPA calculations \cite{lac} and close to the experimental value $E \sim 20 MeV$ \cite{woo}.
 For the $^{40}Ca$ formed by fusion (Fig. $3-a$) it is around 
$150$ fm/c. This large difference is due to the deformation of the
fused system. Indeed, as we can see in Fig. 1 the compound nucleus
 relaxes its initial prolate elongation along the axis of the
collision with a typical time much larger than the dipole oscillation's period. 
 During the considered time window, the averaged value of the observed quadrupole deformation
parameter defined by $\epsilon = \frac{<Q_{20}>}{2 \sqrt{5} <Q_{00}>}$
is around $\epsilon =0.23$. 

\begin{figure}[htbp]
\begin{center}
\hspace*{0.5cm} \epsfig{file=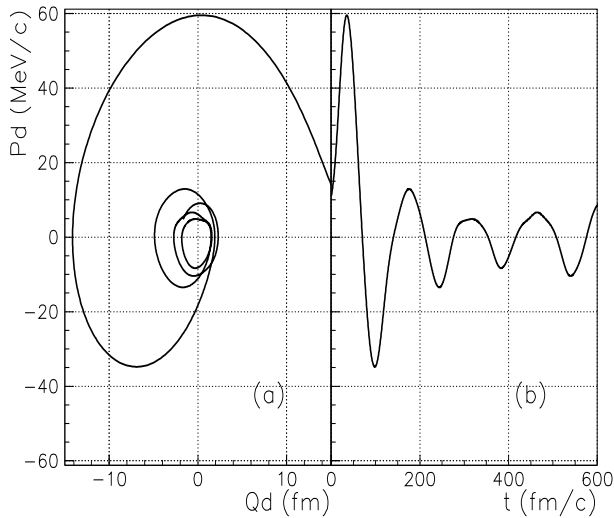,height=7.0cm,width=8.0cm}  
\end{center}
\caption{{{{{{{{\protect\footnotesize {Time evolution of the dipole
vibration. Dipole moment Qd and its conjugate Pd are plotted in the phase
space (a) and Pd is plotted in function of time (b). }}}}}}}}}
\label{fig:spiral}
\end{figure}

A lower mean energy is expected for this longitudinal collective
 motion $E_{GDR_{X}}$ as compared to the one simulated in a spherical $^{40}Ca$. Following
 a macroscopic model for the dipole oscillation we expect the energy of the GDR to evolve 
 with the deformation as 
 \begin{equation}
E_{GDR_{X}}=E_{GDR}\left( 1-\epsilon \right) ^{2}  \label{Eq:Egdr}
\end{equation}
 The frequency of the GDR along the elongation
 axis, fullfill this relation with $\epsilon \approx 0.26$ in good agreement with the observed 
 average deformation.
 
   The dominant role of the deformation in the lowering of the GDR energy 
   can be easily reproduced by performing a Hartree Fock calculation of a $^{40}Ca$ nucleus in 
an external quadrupole field. This external field is kept during all the
dynamical evolution so that the deformation does not relax. At time $0$ a
dipole oscillation is induced by a dipole boost. The resulting dipole
oscillations have a lower oscillation period. Using the deformation parameter $\epsilon =0.23$
we find that this period is close to the one observed in the fusion case and agrees with the phenomenological
relation (\ref{Eq:Egdr}).

\smallskip

\begin{figure}[tbph]
\begin{center}
\hspace*{0.5cm} \epsfig{file=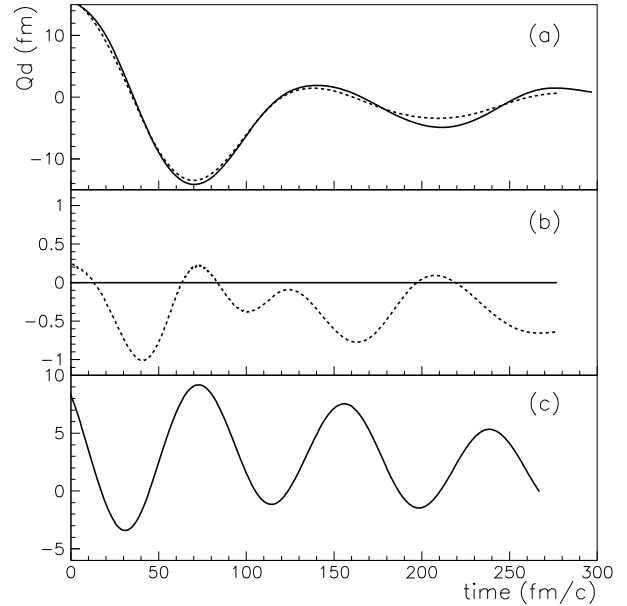,height=8.0cm,width=8.0cm} 
\end{center}
\caption{{{{{{{{\protect\footnotesize { Time evolution of the component of the 
dipole moment parallel to the deformation axis (a) and the component perpendicular
in the collision plane (b)
for the reaction $^{20}O+^{20}Mg$ with the impact parameter $b = 0 fm$ (solid line)
and $b = 4 fm$ (dashed line). (c) GDR in the $^{40}Ca$ with a spherical shape }}}}}}}}}
\label{fig:moment}
\end{figure}

  To get a deeper insight in the dipole oscillation observed
in fusion reactions we have analyzed the time evolution of its period. From
each point on the collective trajectory this quantity can be inferred from
the time needed to reach the opposite side of the observed spiral. The
resulting evolution is plotted in Fig. $4-a$. This period shows
oscillations too. On the other hand, we calculated 
the  monopole moment $Q_{00}$ and the quadrupole moment $Q_{20}$ which are 
presented in figure $4-b$. We can see that those observables  oscillations
which are almost in phase with those in Fig. $4-a$. This points toward a possible coupling between the
dipole mode and another mode of vibration.  The evolutions of the monopole
and quadrupole moments are very similar. In particular, they   have  the
same oscillation's period around $166$ fm/c. Therefore, we conclude that they originates from
the same phenomenon, the vibration of the density around a prolate shape.
This oscillation modifies the properties of the dipole mode in a time
dependent way.

  In order to investigate if this density vibration is the origin 
of the observed behavior of the dipole period we 
can model the induced mode coupling. We consider a harmonic oscillator
with a spring constant which varies in time at a frequency $\omega $. This is a
simple way to take into account the fact that oscillations of the density, which
enters in the mean field potential of the TDHF equation, modify the dipole
restoring force. This is a way to take into account the non linear behaviour of 
mean field dynamic. 

\begin{figure}[htbp]
\begin{center}
\hspace*{0.5cm} \epsfig{file=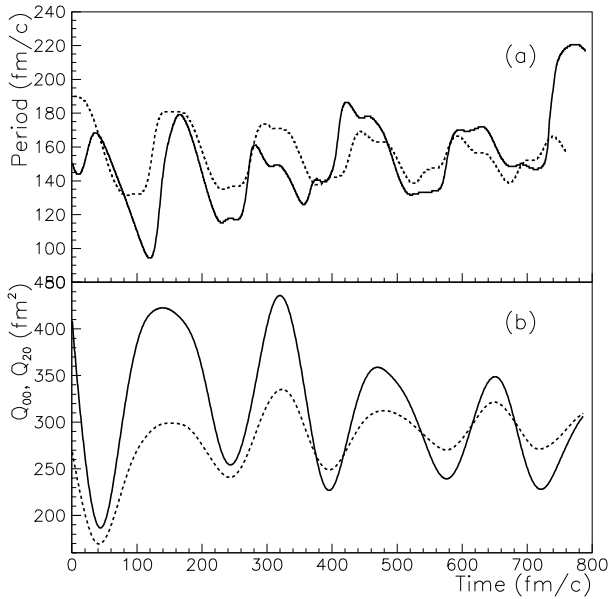,height=8.0cm,width=8.0cm} 
\end{center}
\caption{{{{{{{{\protect\footnotesize {(a) Time behaviour of the dipole
period (solid line) and its modelisation with the Mathieu's equation (dashed
line). (b) Time evolution of the monopole (dashed line) and quadrupole
moments (solid line).}}}}}}}}}
\label{fig:period}
\end{figure}

It is easy to understand that a lower density as well as in an
elongated shapes induce a weaker restoring force. Thus, variations of the density
profiles in the TDHF equation can be modeled by a corresponding variation of
the constant of rigidity in a spring Hamiltonian. In such a model, the equation of
motion becomes the Mathieu's equation 
\begin{equation}
\frac{\ddot{x}}{\omega _{0}^{2}}+[1+\beta (\frac{\omega ^{2}}{\omega
_{0}^{2}}+2\frac{\omega }{\omega _{0}})cos(\omega t)]x=0  \label{Eq:Mathieu}  
\end{equation}
where $\omega _{0}$ is the pulsation without coupling and $\omega $ the
pulsation of the density's oscillation while $\beta $ corresponds to
the magnitude of the induced frequency fluctuations. We have computed the
numeric solution of equation (\ref{Eq:Mathieu}) with the typical external
frequency of the monopole and quadrupole oscillations. The bare frequency $%
\omega _{0}$ and the coupling strength $\beta $ have been tuned in
order to get the same dipole oscillation frequency and typical strength as in the
 full TDHF calculations. Indeed,
because of the presence of the oscillating term, the observed frequency is
different from the bare one. The numerical solution of the Mathieu equation shows oscillations
 with an oscillating period which well reproduces our self consistent calculations
with $\beta =0.15$ and $\omega _{0}$ changed by a factor 1.2 from the
observed value $E_{GDR_{X}}$ (see Fig. 4-a). From this analysis it appears
 that the observed dipole motion corresponds to a giant vibration along the main axis
of an oscillating prolate shape. The observed modulation of the dipole
frequency is a source of additionnal spreading of the resonance line
shape.

Using coherent state picture the maximal elongation $d_{max}=\frac{A}{N Z} Q_{d_{max}}
\approx 1.4fm$ is related to the
number of excited phonons by the expression 
\begin{equation}
n_{0}=d_{max}^{2}\frac{ME_{GDR_X}}{2\hbar ^{2}}   \label{Eq:Phonon}  
\end{equation}
where $E_{GDR_X}\approx 8$MeV is the energy of the GDR along the $(x)$ axis and 
$M=\frac{NZ}{A} m$ is the reduced mass of the neutron-proton system, 
m being the nucleon mass. It gives $n_{0}\approx %
1.9$. Obviously this value is an upper limit because the
mean field dynamics underestimate the consequences of the damping since 
two body dissipation is not taken into account. In the
same way as ref. \cite{cho2,fli,cho1,bor}, we can compute the $\gamma $%
-decay probability $P_{\gamma }$ over $\overline{P_{\gamma }}$ its mean
value in the equilibrium with the expression
\begin{equation}
\frac{P_{\gamma }}{\overline{P_{\gamma }}}=\frac{\Gamma ^{\downarrow }+\frac{%
n_{0}}{\overline{n_{GDR}}}\Gamma _{evap}}{\Gamma ^{\downarrow }+\Gamma
_{evap}}
\end{equation}
where $\Gamma _{evap}$ is the rate at which the compound system  decay (see
ref. \cite{cho1}), $\Gamma ^{\downarrow }$ is the decaying rate of the
phonons and $\overline{n_{GDR}}$ is the mean number of excited phonons in
the equilibrium. The latter can be estimated by  $\overline{n_{GDR}} = 3 e^{\frac{-E_{GDR}}{T}}$
where $T$ is the compound nucleus temperature. For $^{40}Ca$ at an excitation energy around 96 
MeV, we take $\Gamma _{evap}\approx 0.76MeV$, $\Gamma ^{\downarrow }%
\approx 7MeV$ and $\overline{n_{GDR}}\approx 0.6$. It gives $\frac{%
P_{\gamma }}{\overline{P_{\gamma }}}\approx 1.2$ which shows that the
pre-equilibrium effects are only a correction for this system since the value for an entrance channel 
quenching  ($n_0 = 0$) of the GDR using the isospin symmetric reaction $^{20}Ne+^{20}Ne$ would 
lead to $\frac{P_{\gamma }}{\overline{P_{\gamma }}}\approx 0.9$.
 It means the first chance gamma emission will be slightly enhanced by about a factor 20 \%.
 This is in good agreement with the values reported in the litterature \cite{fli}.

 In order to compute the properties of the pre-equilibrium GDR in fusion we shall 
  also study non zero impact parameters. In particular the interplay of dipole vibration 
and deformation can be affected by the rotation. In addition to the center of mass 
coordinate system with $x$ along the beam axis and $y$ perpendicular to the reaction plane
we defined a new coordinate system ${x',y',z'}$ where  $x'$ is the deformation axis
 and $y=y'$ is the rotation axis. For the head on head reaction
studied before, those two frames are the same. In this case, for symmetry reasons the dipole moments along the 
$z'=z$ and $y$ axis are zero at every time.

If we consider now a finite $b$, the symmetry only forbids the vibration to  occur along $y$.
 In this case, the amplitude of the oscillations along $x'$ slightly decreases with the impact parameter 
(Fig. $3-a$, dashed line) whereas an oscillation along $z'$ appears (Fig. $3-b$, dashed line).
First we can see that the period of the dipole oscillation perpendicular to the deformation axis
 is lower than one along the deformation axis.
It is another consequence for the prolate deformation of the nucleus indeed the energy 
of this excitation along $z'$ should obey the relation $E_{GDR_{Z'}} = E_{GDR} (1+\epsilon)$.
This relation gives a value of the oscillation's period of 67 fm/c which is the
observed one on (Fig $3-b$).
The oscillation along $z'$ shows a lower amplitude than this along $x'$ 
(approximatively 1 per cent in relative intensity). This oscillation along the $z'$ axis
results from a weak symmetry breaking due to the rotation of the system. Indeed the changing to 
the rotating frame $x',y=y',z'$  leads to the Hamiltonian with a non diagonal term
\begin{equation}
H' = R(t) H R^{-1}(t) + \dot{\alpha} J_y
\end{equation}
where $H'$ is the Hamiltonian in the rotating frame, $\dot{\alpha}$ is 
time derivative of the angle between the two frames and $J_y$ is 
the generator of the rotations around $y$.
The last term induces a motion along the $z'$ axis from a dipole vibration along $x'$. This is indeed
what is happening since initially the dipole mode is along the $x'$ axis. 

If we now compute the number of excited phonons we see almost 
no variation of the total phonon number $n_0$ with $b$. Therefore the 
conclusions reached for $b=0$ can be extended to the full range of impact parameters $b$. 
 
  Let us now study the role of the incident energy. The lack of two body damping in TDHF
doesn't enable us to go up in energy with this system. So we investigate 
a lower energy 0.5 MeV per nucleon
which is the coulomb barrier. In this case, the number of exciting phonons is 1.0.
This value is lower than one obtained at 1MeV/u because the maximum dipole moment decreases with decreasing energy.
Therefore the number of pre-equilibrium phonons is lower at lower energy. 
 
 We have also investigated the role of the mass of the reactions partners to study if the pre-equilibrium effects 
 survive in heavier compound systems. We have computed the $^{40}Ar+^{40}Ti$ at 1 MeV per nucleon
 in the center of mass system (slightly above the fusion barrier).
 We have observed exactly the same phenomenology with a strong excitation of pre-equilibrium GDR. 
 The calculated $n_0 = 0.6$ is smaller than the $n_0$ obtained for lighter nuclei
 reaction. This comes from the fact that the reaction with small nuclei is less damped than the one involving
 more nucleons. However the pre-equilibrium GDR is also strong in the heavier fusion reaction.

  In summary, pre-equilibrium effects related to the N/Z
asymmetry in the entrance channel have been for the first time investigated
using the TDHF approach. The relation between the mean oscillation's period and the shape 
of the nucleus have been established. Because of the deformation the average
dipole frequency along the deformation axis is lower than the usual GDR one ,while
 we observe a weak vibration at higher frequency in the perpendicular direction 
 in the collision plane. In this direction, the oscillation occurs
 only for a non zero impact parameter. The evolution of the
period have been modeled by a Mathieu's spring model so that the link
between the observed phenomena and the giant dipole resonance have been
established. The frequency modulation due to this non-linear coupling between modes
 plays the role of an additionnal spreading of
the GDR frequency. The results shows that pre-equilibrium gamma
emission are expected in the case of fusion reaction with asymmetric N/Z
nuclei. 

  We want to thank Herve Moutarde for his help about the Ma\-thieu's equation and 
Paul Bonche for providing his TDHF code and for helpful discussion at the initial stage of this work.
 Useful discussions with Pieter Van Isacker and Denis Lacroix are acknowledged.

\end{document}